\documentclass[letterpaper]{jpconf}
\usepackage{graphicx}
\begin{document}
\title{The Evolution of Correlation Functions from Low To High $p_T$ in PHENIX: From HBT To Jets}

\author{J.T. Mitchell and the PHENIX Collaboration}

\address{Brookhaven National Laboratory, P.O. Box 5000, Building 510C, Upton, NY 11973-5000}

\ead{mitchell@bnl.gov}

\begin{abstract}
The PHENIX Experiment at the Relativistic Heavy Ion Collider has performed a survey of momentum correlations ranging from 200 MeV/c to 7 GeV/c in $\sqrt{s_{NN}}$ = 200 GeV p+p, d+Au, Au+Au, and $\sqrt{s_{NN}}$ = 62 GeV Au+Au collisions. The correlations are measured separately for like-sign and unlike-sign pairs. Comparisons of the properties of the near-side peak amplitude and width as a function of centrality and transverse momentum for each collision species are presented and discussed.
\end{abstract}.

\section{Introduction}

The technique of two-particle correlations has demonstrated its effectiveness in extracting information about jet production and the interaction of jets with the medium at high transverse momentum in reference (p+p and d+Au) and heavy-ion collisions. The data suggest that the correlations are dependent on the $p_T$ of the trigger and associated particles. Therefore, it is also important to examine and understand the evolution of correlation functions from high to low $p_T$ and look for differences in that evolution between reference and heavy ion collisions. Using this method, any features such as the possible presence of excess jet production at low $p_T$ can be addressed.

Interpreting correlation functions at low $p_T$ (below 1 GeV/c) is complicated by several factors: a) the hard scattering back-to-back correlation becomes reduced in amplitude and diluted in azimuthal separation, b) correlations due to the HBT effect are present, c) correlations due to resonance production are present, and d) correlations due to collective flow are present. In order to help disentangle the latter two contributions, all correlations are separated into only like-sign or only unlike-sign pairs. Due to space limitations, only the like-sign results, which should not have a significant resonance contribution, are presented here.

Details about the PHENIX experimental configuration can be found elsewhere \cite{phenixNIM}. All of the measurements described here utilized the PHENIX central arm detectors. The minimum bias datasets analyzed are 10 million events from the RHIC Run-3 p+p run, 10 million events from the Run-3 d+Au run, 2 million events from the Run-4 200 GeV Au+Au run, and 750,000 events from the Run-4 62 GeV Au+Au run.  All correlation functions are constructed from a strict mixed event procedure with all cuts, including track proximity cuts, applied identically to the data and mixed event samples. Mixed events are constructed from identical data classes (distinguished by centralities within 5\% and vertex z coordinates within 10 cm) by directly sampling the data charged particle multiplicity distribution after all cuts have been applied.

\section{Data and Discussion}

With the majority of events containing trigger particles in the $p_T$ range from 0.2 to 7.0 GeV/c, the correlations were constructed by considering every particle in the event as a trigger particle in succession and pairing it with every other particle in the event. Correlations constructed in this manner have been referred to as auto-correlations \cite{starAuto}. The correlation amplitudes reported here are normalized in the following manner: $C = \frac{N_{data}/N_{events,data}}{N_{mixed}/N_{events,mixed}}$, where $N_{data}$ and $N_{mixed}$ are the number of pair counts in a given $p_T$ and/or $\Delta \phi$ bin.

\begin{figure}[h]
\begin{minipage}{18pc}
\includegraphics[width=18pc]{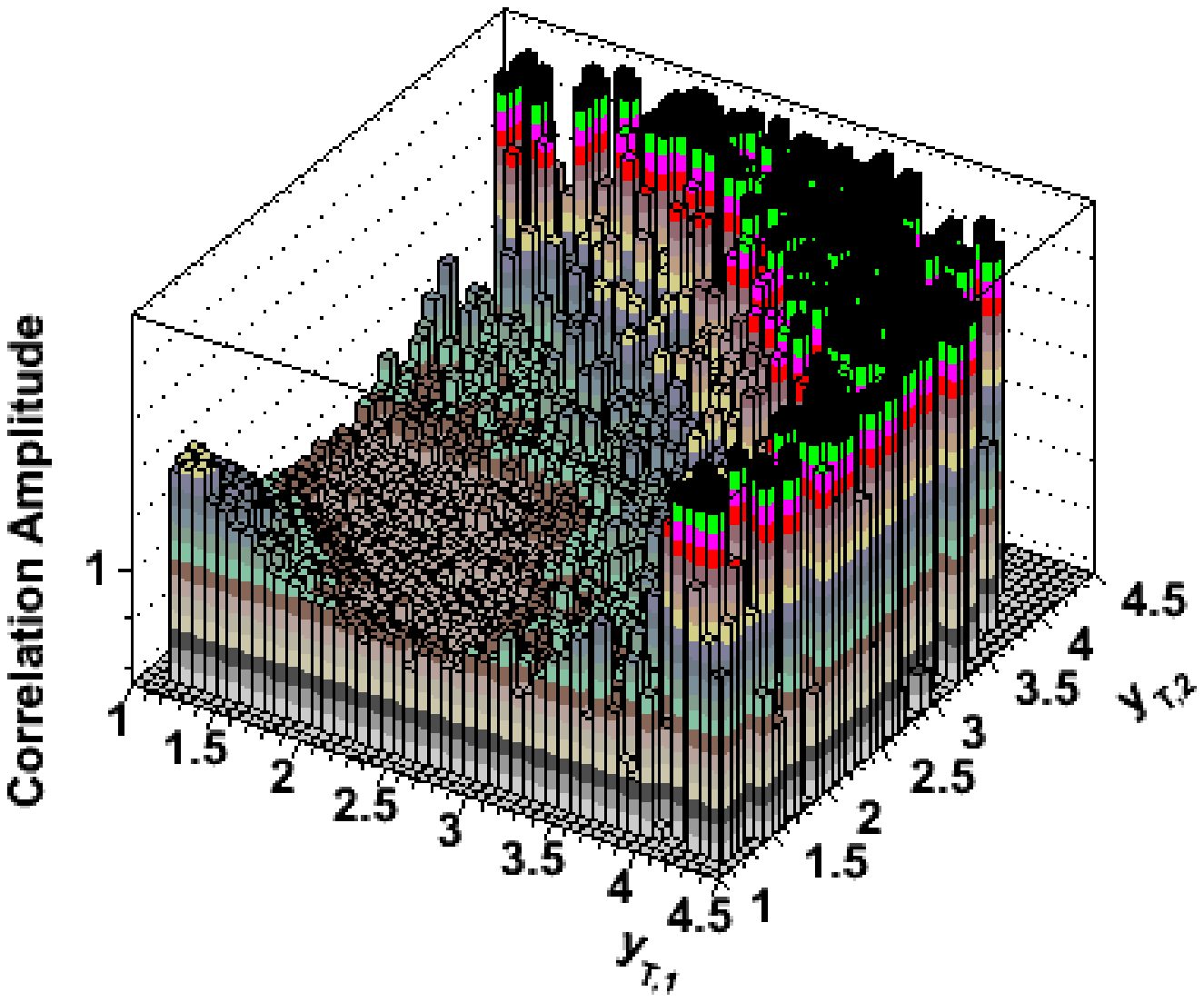}
\caption{\label{fig:dauytLSNear}Near-side correlation amplitudes for like-sign pairs as a function of the transverse rapidity of each particle in the pair for minimum bias 200 GeV d+Au collisions.}
\end{minipage}\hspace{2pc}%
\begin{minipage}{18pc}
\includegraphics[width=18pc]{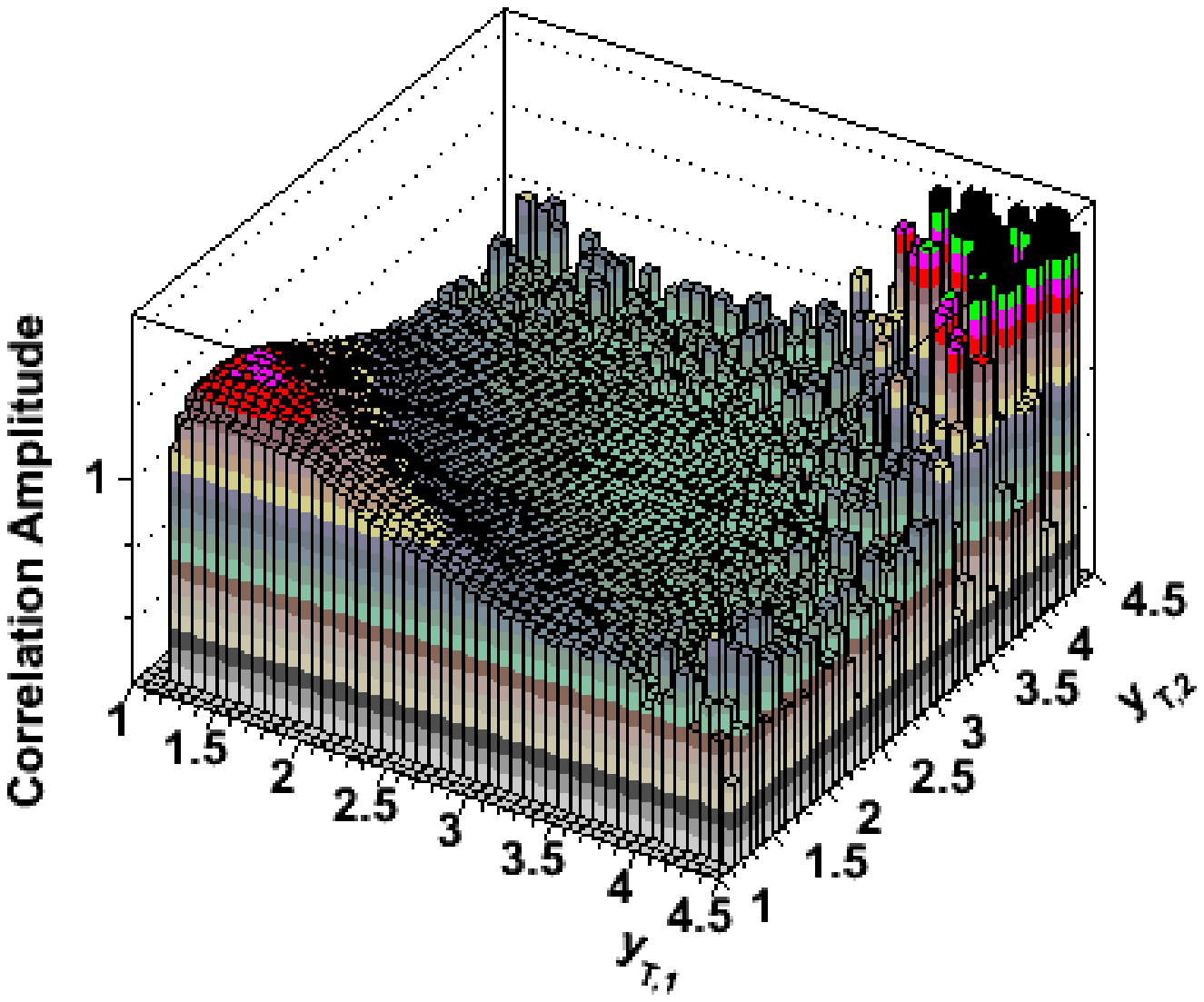}
\caption{\label{fig:auauytLSNear}Near-side correlation amplitudes for like-sign pairs as a function of the transverse rapidity of each particle in the pair for 0-5\% central 200 GeV Au+Au collisions.}
\end{minipage}\hspace{2pc}%
\end{figure}

The correlation amplitudes plotted as a function of the transverse rapidity, $y_T = ln((m_T + p_T)/m_0)$ with the pion mass assumed for $m_{0}$, for like-sign pairs on the near-side, defined as particle pairs with $|\Delta \phi|<60^{o}$, are shown in Fig. \ref{fig:dauytLSNear} for minimum bias 200 GeV d+Au collisions and Fig. \ref{fig:auauytLSNear} for 0-5\% central 200 GeV Au+Au collisions.  There is no attempt to subtract a baseline correlation from the extracted amplitudes.  Instead, direct comparisons to the reference and Au+Au results will be made. The $y_T$ variable is chosen in order to emphasize the distribution at low $p_T$ with respect to that at high $p_T$ and to facilitate comparisons to measurements by the STAR Collaboration \cite{starAuto}. For reference, $y_T=1.5$ corresponds to $p_T=300$ MeV/c and $y_T=2.7$ corresponds to $p_T=1.0$ GeV/c for pions. There are two primary features observed in these plots.  First, there is the expected large correlation amplitude mostly due to hard scattering processes when either particle in the pair has a large $p_T$. Second, there is a peak when both pairs have low $p_T$. This peak has a large contribution from correlations due to HBT, which has been confirmed by observing a sharp reduction in the peak amplitude for unlike-sign pairs and by observing a peak in the $Q_{invariant}$ distribution in this region for like-sign pairs. Examination of this peak reveals that there is a difference in the location of the peak in $y_T$ between d+Au and Au+Au collisions. In d+Au collisions, the peak amplitude rises up to the low $p_T$ edge of the detector acceptance. This holds for p+p collisions, also. However, there is a maximum in the peak at $y_T \approx 1.4$ ($p_T \approx 250$ MeV/c) in Au+Au collisions at both 200 and 62 GeV, which warrants a more detailed study of this effect.

\begin{figure}[h]
\begin{minipage}{18pc}
\includegraphics[width=18pc]{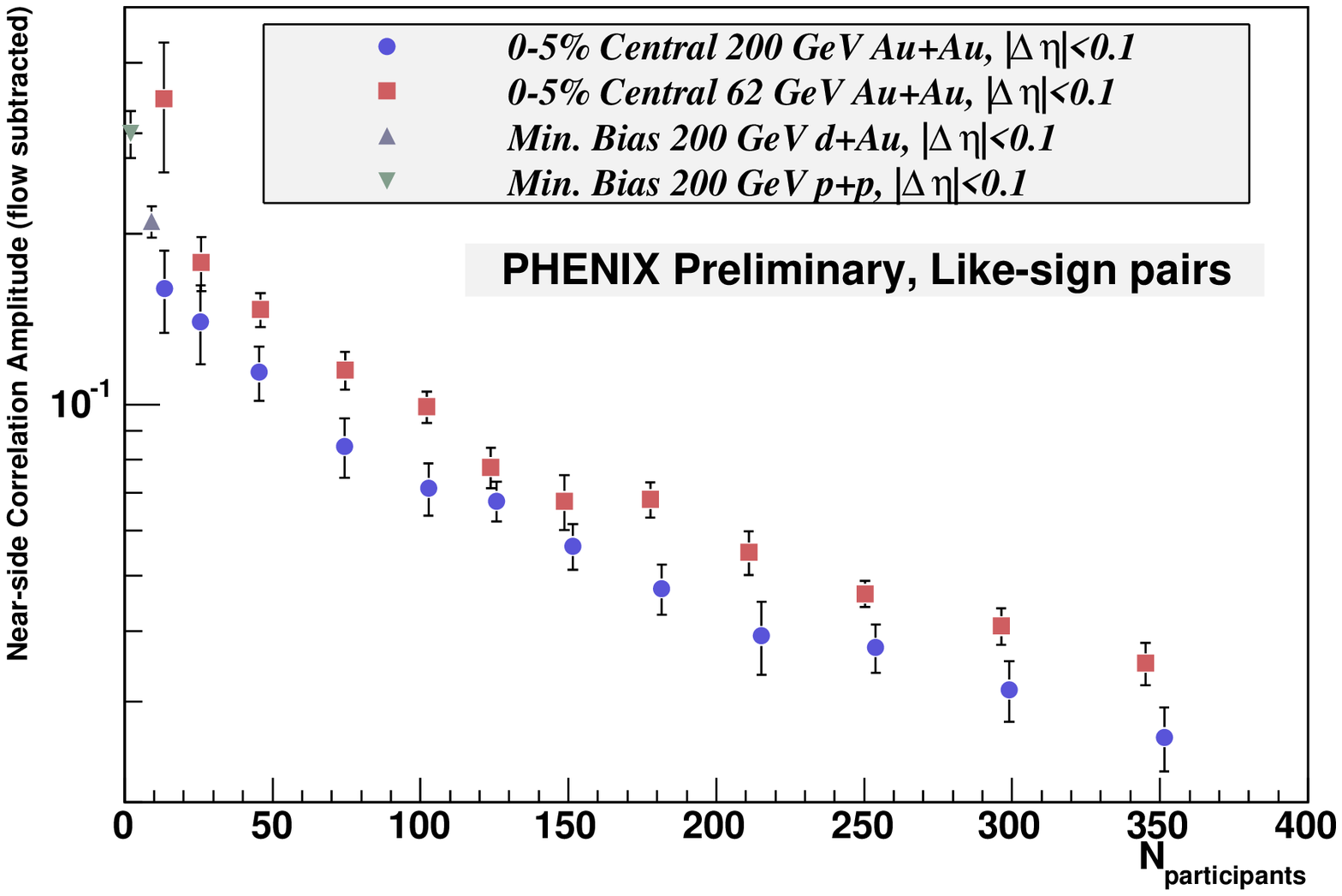}
\caption{\label{fig:ampVsCentLSNS}Near-side correlation amplitudes for like-sign pairs with $p_T$ between 200 and 500 MeV/c and $|\Delta \eta|<0.1$ for p+p, d+Au, and Au+Au collisions. Flow has been subtracted from the Au+Au data. The errors are dominated by systematic errors due to the fit procedure.}
\end{minipage}\hspace{2pc}%
\begin{minipage}{18pc}
\includegraphics[width=18pc]{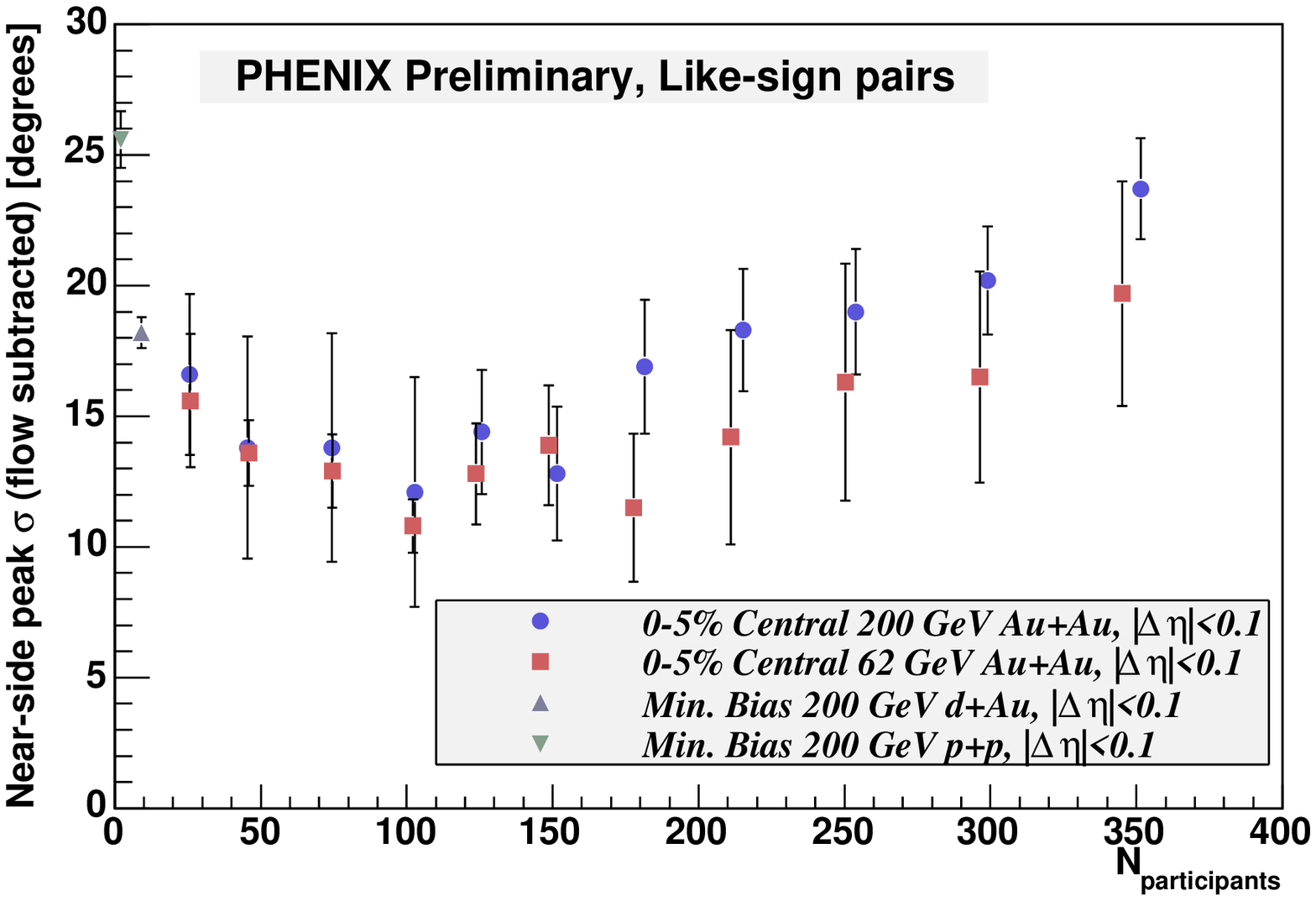}
\caption{\label{fig:sigmaVsCentLSNS}Near-side correlation widths for like-sign pairs with $p_T$ between 200 and 500 MeV/c and $|\Delta \eta|<0.1$ for p+p, d+Au, and Au+Au collisions. Flow has been subtracted from the Au+Au data. The errors are dominated by systematic errors due to the fit procedure.}
\end{minipage}\hspace{2pc}%
\begin{minipage}{18pc}
\includegraphics[width=18pc]{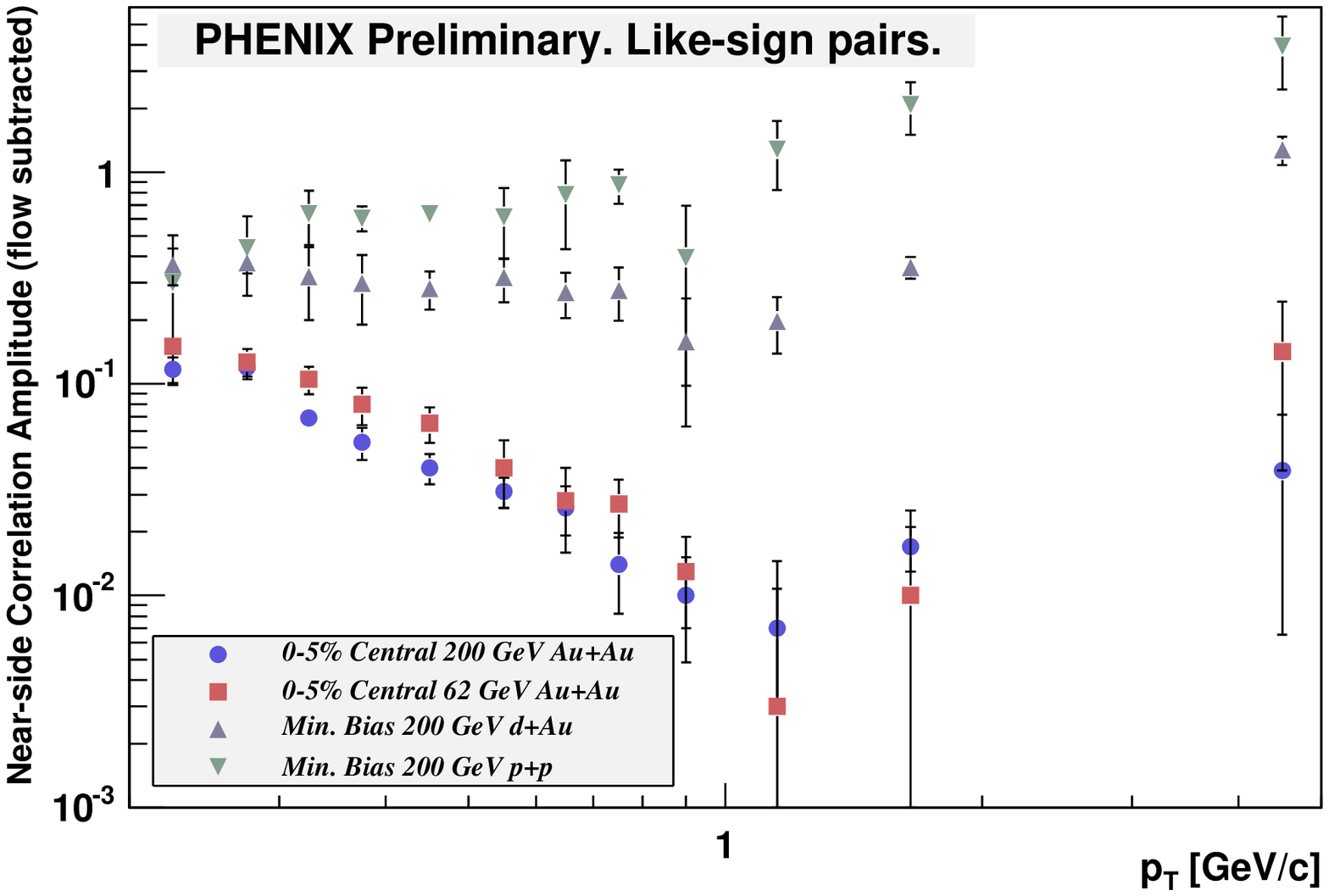}
\caption{\label{fig:ampVsPtLSNS}Near-side correlation amplitudes for like-sign pairs both with $p_T$ within the given $p_T$ bin and with $|\Delta \eta|<0.1$ for p+p, d+Au, and 0-5\% central Au+Au collisions. Flow has been subtracted from the Au+Au data. The errors are dominated by systematic errors due to the fit procedure.}
\end{minipage}\hspace{2pc}%
\end{figure}

In Fig. \ref{fig:auauytLSNear}, flow has not been subtracted, so there is a flow contribution to the higher $p_T$ regions of this plot. However, at low $p_T$, the flow component is very small and does not contribute significantly to the peak there. In order to directly compare the d+Au and Au+Au correlations for the entire $p_T$ range, it is necessary to subtract the flow component from the Au+Au azimuthal correlation distributions. This is done by fitting the near-side region ($\Delta \phi < 90^{o}$) of these distributions with a function containing a harmonic $cos(2 \Delta \phi)$ term and a Gaussian distribution with a mean of $\Delta \phi = 0$. For p+p and d+Au collisions, the Gaussian fit is performed, but the harmonic term is not included. 

Fig. \ref{fig:ampVsCentLSNS} shows the correlation amplitude from the near-side Gaussian fit for like-sign pairs as a function of centrality. For this plot, all pairs have a $p_T$ between 200 and 500 MeV/c. In order to better select the HBT contribution to the peak, all pairs are required to be within $|\Delta \eta|<0.1$. For both 200 and 62 GeV Au+Au collisions, the amplitude drops exponentially with $N_{participants}$, as expected from random dilution of the correlation amplitude \cite{vol04}. The difference between the 200 and 62 GeV amplitudes is also largely due to the effect of random dilution.

Fig. \ref{fig:sigmaVsCentLSNS} shows the standard deviation of the near-side Gaussian fit for like-sign pairs as a function of centrality. Again, all pairs have a $p_T$ between 200 and 500 MeV/c and $|\Delta \eta|<0.1$. The error bars are dominated by systematic errors due to the fitting procedure. There is a large difference observed in the width between p+p and d+Au collisions. In Au+Au collisions, particularly for the higher statistics 200 GeV Au+Au dataset, a significant increase in the width is seen in the most central collisions. 

Fig. \ref{fig:ampVsPtLSNS} shows the correlation amplitude for the near-side Gaussian fit for like-sign pairs as a function of transverse momentum. Here, $p_{T,min}<p_{T,1}<p_{T,max}$ and $p_{T,min}<p_{T,2}<p_{T,max}$ ($p_T$ within the limits of the bin, $p_{T,min}$ and $p_{T,max}$) and $|\Delta \eta|<0.1$. The points are plotted in the geometric center of the bin. The bins are exponentially distributed in order to nullify any variations due to random dilution of the correlations within each collision species. In both the p+p and d+Au datasets, the amplitudes are relatively flat as a function of $p_T$ up to about 1-1.5 GeV/c, where the amplitudes begin to rise significantly. This rise is likely due to the influence of correlations due to jets, which has been confirmed by the appearance of an away-side component in the azimuthal correlations. A rise in the correlation amplitudes for Au+Au collisions is also seen starting at the same $p_T$. However, unlike the p+p and d+Au datasets, the Au+Au data exhibit an exponential decrease in the amplitude as a function of $p_T$ below 1 GeV/c. This decrease is contrary to the trend in the reference measurement and to the ansatz of significant contributions from jet production at low $p_T$. More studies of the cause of this effect are underway.

\section{Summary}

PHENIX has presented two-particle correlation amplitudes for like-sign near-side pairs as a function of centrality and transverse momentum. Differences between the baseline p+p and d+Au collisions and Au+Au collisions are observed in the correlation amplitudes as a function of $p_T$.

\smallskip

\end{document}